\documentclass[12pt,a4paper]{article}

\usepackage[english]{babel}
\usepackage{psfrag,graphicx}
\usepackage{amsmath,amssymb,bbm,stmaryrd}
\usepackage{hyperref}

\newcommand{\dx}{{\rm d}}
\newcommand{\Id}{\mathbbm{1}}
\newcommand{\E}{\mathbbm{E}}
\newcommand{\Pb}{\mathbbm{P}}
\newcommand{\Or}{\mathcal{O}}
\newcommand{\R}{\mathbbm{R}}
\newcommand{\Z}{\mathbbm{Z}}
\newcommand{\N}{\mathbbm{N}}
\newcommand{\Perm}{\mathcal{S}}

\DeclareMathOperator*{\Airy}{\cal A}
\DeclareMathOperator*{\Ai}{Ai}
\DeclareMathOperator*{\Aip}{Ai^\prime}
\DeclareMathOperator*{\Var}{Var}
\DeclareMathOperator*{\Pf}{Pf}
\DeclareMathOperator*{\Tr}{Tr}

\numberwithin{equation}{section}

\newenvironment{changemargin}[2]{\begin{list}{}{%
\setlength{\topsep}{0pt}%
\setlength{\leftmargin}{0pt}%
\setlength{\rightmargin}{0pt}%
\setlength{\listparindent}{\parindent}%
\setlength{\itemindent}{\parindent}%
\setlength{\parsep}{0pt plus 1pt}%
\addtolength{\leftmargin}{#1}%
\addtolength{\rightmargin}{#2}%
}\item }{\end{list}}
\title{One-dimensional stochastic growth and Gaussian ensembles of random matrices}
\author{Patrik L. Ferrari and Michael Pr\"ahofer\\ {\normalsize Zentrum Mathematik} \\ {\normalsize Technische Universit\"at M\"unchen} \\{\normalsize e-mails: ferrari@ma.tum.de, praehofer@ma.tum.de}}
\date{12th August 2005}

\begin{document}
\maketitle

\begin{abstract}
In this review paper we consider the polynuclear growth (PNG) model in one spatial dimension and its relation to random matrix ensembles. For curved and flat growth the scaling functions of the surface fluctuations coincide with limit distribution functions coming from certain Gaussian ensembles of random matrices. This connection can be explained via point processes associated to the PNG model and the random matrices ensemble by an extension to the multilayer PNG and multi-matrix models, respectively. We also explain other models which are equivalent to the PNG model: directed polymers, the longest increasing subsequence problem, Young tableaux, a directed percolation model, kink-antikink gas, and Hammersley process.
\end{abstract}

%{\footnotesize {\noindent {\sc Keywords:} KPZ growth, PNG model, Young tableaux, Gaussian random matrices, determinantal and Pfaffian point processes\\
%{\sc AMS Subject Classification:} Primary 82C05, 15A52; Secondary 60G55\\
%{\sc Running title:} Polynuclear growth model and random matrices}}

\section{Introduction}\label{Intro}
In this paper we consider a stochastic growth model, the polynuclear growth (PNG) model, on a one-dimensional substrate. The relevant features of the dynamics of this model are: a \emph{stochastic local growth rule} with a \emph{smoothing mechanism}~\cite{PV98}. The latter prevents the formation of large spikes. As a consequence the surface, on a macroscopic scale, follows a deterministic growth rule: it has a limit shape. Nevertheless, on a mesoscopic scale the surface is still rough. This roughness is the observable we are mainly interested in. The PNG model belongs to the KPZ universality class. The KPZ model was introduced by Kardar, Parisi, and Zhang in a seminal paper~\cite{KPZ86} where they described random surface growth by a non-linear stochastic differential equation. For details on the universality we refer to Pr\"ahofer's thesis~\cite{Pra03}, Chapters 2 and 3, see also~\cite{PS02b}. In one dimension one can determine the dynamical exponent, $z=3/2$, by a scaling argument or renormalization group methods, see for example the books~\cite{PV98,BS95}. As a consequence, for large growth time $t$, the fluctuations of the surface height scale as $t^{1/3}$ and the spatial correlation length is of order $t^{2/3}$. These exponents should hold for all growth models in the KPZ universality class. It is commonly expected that not only exponents but also scaling functions and limiting distributions are universal. To study these more detailed informations on KPZ growth, one looks for simplified but still solvable models in the KPZ class. Hence the study of the PNG model.

A main step forward in understanding KPZ growth and suggesting a deep connection to random matrices was achieved by Johansson~\cite{Jo00b}. He considered a discrete growth model on a one-dimensional substrate, which can be interpreted, among others, as a kind of first-passage site percolation model~\cite{Dha87}. For specific initial conditions, he obtained the surprising link between the shape fluctuations of the percolated region and the GUE Tracy-Widom distribution, $F_2$, which was first introduced in the random matrix context~\cite{TW94}. Shortly before that, the same distribution appeared in a work by Baik, Deift, and Johansson~\cite{BDJ99} on the problem of the longest increasing subsequence in a random permutation. On the other hand there was the PNG model, a model considered by physicists since the 60's. It was known that it belongs to the KPZ class. Then Pr\"ahofer and Spohn~\cite{PS00} noticed the mapping between (a version of) the PNG model and the increasing subsequences problem. Hence the connection between KPZ and random matrices.

Here we mainly focus on two geometries of the PNG model: curved growth and flat growth. The first generates a droplet-like profile, hence called \emph{PNG droplet}. The macroscopic profile of the second is flat, thus \emph{flat PNG}. For these two geometries more refined information is available. In the limit of large growth time, properly rescaled, the height fluctuations of the PNG droplet are described by the $F_2$ distribution~\cite{PS00,BDJ99}.  $F_2$ is the limiting distribution of the properly rescaled largest eigenvalue of the Gaussian Unitary Ensemble (GUE) of random matrices. Moreover, the limiting process describing the surface height of the PNG droplet has been identified as the Airy process~\cite{PS02}, which also arises in the multi-matrix extension of GUE. It describes the evolution of the largest eigenvalue of Dyson's Brownian motion.

For the second geometry considered, the flat PNG, the picture is not yet complete. It is known~\cite{PS00,BR99} that the scaling function of the height fluctuation is the GOE Tracy-Widom distribution, $F_1$~\cite{TW96}. This distribution arises in the Gaussian Orthogonal Ensemble (GOE) of random matrices. $F_1$ is the limiting distribution of the largest eigenvalue of GOE. It is conjectured that it should correspond to the evolution of the largest eigenvalue of Dyson's Brownian motion (for orthogonal matrices). A result going in the direction of this conjecture is obtained by Ferrari~\cite{Fer04}. The problem remains open for the GOE multi-matrix model. Very recently, Sasamoto~\cite{Sas05} obtained a process in the context of the totally asymmetric exclusion process, which, by universality, should describe also the surface height of the flat PNG.

The connection between the PNG model and the Gaussian ensembles of random matrices can be understood via point processes. Although random matrices are not directly related to the PNG model, it turns out that both can be described by point processes with the same mathematical structure. Under an appropriate scaling one obtains the same limit point processes when the growth time, resp.\ matrix dimension, tends to infinity. 

The paper is organized as follows. In Section~\ref{PNGmodel} we introduce the PNG model and report known results. In Section~\ref{OtherModels} further equivalent models are described. In Section~\ref{LineEnsembles} we explain the extension to the PNG multilayer model and the related point processes. The link between line ensembles and (real-valued) Young tableaux is also discussed. In Section~\ref{RandomMatrices} we introduce Gaussian ensembles of random matrices, the point processes of their eigenvalues, and the extension to multi-matrix models. Section~\ref{Connections} is devoted to the discussion of the connection between random matrices and the PNG model as well as Young tableaux. 

\section{The polynuclear growth model}\label{PNGmodel}
\begin{figure}[t!]
\begin{center}
\psfrag{x}[c]{$x$}
\psfrag{h}[l]{$h(x,t)$}
\includegraphics[height=5cm]{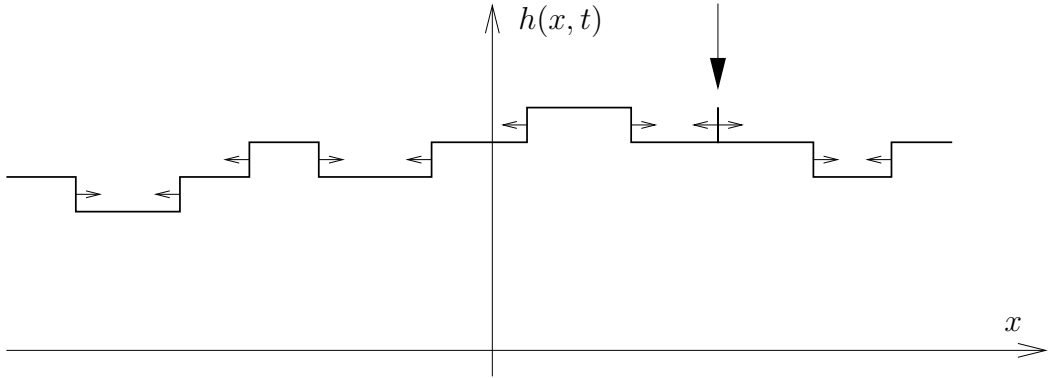}
\caption{The PNG dynamics. The up- (down-)steps move to the left (right) with unit speed. The big arrow represents a nucleation.}\label{FigPNGdynamics}
\end{center}
\end{figure}

Let us describe the polynuclear growth (PNG) model in continuous time in $1+1$ dimension. It is a growth model on a one-dimensional substrate. The surface at time $t$ is described by an integer-valued height function $x\mapsto h(x,t)\in\Z$, see Figure~\ref{FigPNGdynamics}. To be precise, at the discontinuity points of $h$, the height function has upper limits, i.e., $\{x\in \R | h(x,t)\geq k\}$ is a closed set for all $k\in\Z$. For fixed time $t\in \R$, consider the height profile $x\mapsto h(x,t)$. The height $h$, as $x$ increases, has jumps of height one at the discontinuity points, called \emph{up-step} if $h$ increases and \emph{down-step} if $h$ decreases. Finally, if at time $t$ there is a spike, i.e., a pair of up- and down-steps at the same position $x$, then we call it a \emph{nucleation event} at $(x,t)$.

The PNG dynamics has a deterministic and a stochastic part:
\begin{itemize}
\item[(a)] Deterministic part: the up-steps move to the left with unit speed and the down-steps to the right with unit speed. When a pair of up- and down-steps collide, they disappear.
\item[(b)] Stochastic part: the nucleation events form a locally finite point process in space-time (usually a Poisson process). Once a pair of up-down steps is created, it immediately follows the deterministic dynamics.
\end{itemize}
The stochastic part of the dynamics produces the roughness of the surface. This is counterbalanced by the smoothing due to the deterministic part.

Typically one considers flat initial conditions, $h(x,0)=0$ for all $x\in\R$, with the nucleation events given by a Poisson process (not necessarily with uniform intensity). By varying the space-time intensity $\varrho(x,t)$ of the Poisson process, different geometries can be obtained. Below we consider two cases of particular interest. For the \emph{PNG droplet} the nucleations occur with constant intensity in the region spreading with unit speed from the origin in both directions. For the \emph{flat PNG} the nucleations have constant intensity everywhere, thus the surface is statistically translation-invariant. A visualization of these two geometries is provided at \cite{FerRSK}  as a \href{http://www-m5.ma.tum.de/pers/ferrari/homepage/animations/}{Java applet}.

At this point it is convenient to introduce some vocabulary taken over from special relativity. The steps move with speed of light $\pm c$, $c=1$. So their trajectories in space-time have slope $\pm 1$, called \emph{light-like}. The \emph{forward light cone} of a point $(x,t)$ is the set of points $\{(x',t') | \, |x-x'|\leq t'-t\}$, and the \emph{backward light cone} of $(x,t)$ is $\{(x',t') | \, |x-x'|\leq t-t'\}$. With light cone we denote the union of the forward and the backward cone. A path is called \emph{time-like} if it is included in the light cone of each of its points. A path is called \emph{space-like} if the light cone attached to each of its points does not contain any other point of the path.

\subsection{The PNG droplet}
\begin{figure}[t!]
\begin{center}
\psfrag{h}[l]{$h(x,t)$}
\psfrag{x}[c]{$x$}
\includegraphics[height=5cm]{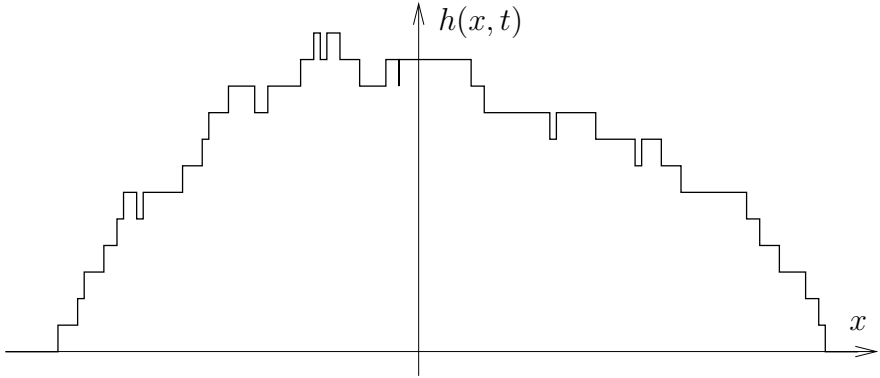}
\caption{A sample of the PNG droplet.}\label{FigDroplet}
\end{center}
\end{figure}

The PNG droplet is obtained from a flat initial height profile, $h(x,0)=0$ for all $x\in\R$, if the density of Poisson points is constant (here we choose $\varrho=2$) in the forward light cone of the origin and zero outside, i.e., for $(x,t)\in\R\times\R_+$
\begin{equation}
\varrho(x,t)=\left\{\begin{array}{ll}
2&\textrm{if } |x|\leq t, \\0&\textrm{if }|x|> t.
\end{array} \right.
\end{equation}
For large growth time $t$ the typical shape of the PNG droplet is a half circle, see Figure~\ref{FigDroplet}.
More precisely, it converges to the limit shape given by
\begin{equation}
\lim_{t\to\infty}t^{-1} h(\tau t,t)=2\sqrt{1-\tau^2},\quad \tau \in [-1,1],
\end{equation}
in probability.

To see the roughness of the surface one has to look at a mesoscopic scale around the macroscopic shape $2t\sqrt{1-(x/t)^2}$, $x\in [-t,t]$. A first natural question concerns the scale of fluctuations and their limit behavior. The result is that the vertical fluctuations live on a $t^{1/3}$ scale with limiting distribution function
\begin{equation}\label{2.3}
\lim_{t\to\infty}\Pb\big(h(0,t) \leq 2t+t^{1/3} s\big) = F_2(s),
\end{equation}
where $F_2$ is the GUE Tracy-Widom distribution function~\cite{TW94}. The convergence is in distribution as well as for all finite moments. (\ref{2.3}) was obtained in~\cite{PS00} by mapping the PNG droplet to the Poissonized version of the problem of longest increasing subsequences~\cite{BDJ99}. Similarly if one looks away from $x=0$ one has, for any fixed $\tau\in(-1,1)$,
\begin{equation}
\lim_{t\to\infty}\Pb\big(h(\tau t,t) \leq 2t\sqrt{1-\tau^2}+t^{1/3}(1-\tau^2)^{1/6}s\big) = F_2(s).
\end{equation}

The second interesting question concerns the spatial height correlations. The correlation length scales as $t^{2/3}$ for large growth time $t$. Therefore one defines the rescaled surface height as
\begin{equation}\label{eq2.8}
\xi\mapsto h^{\rm resc}_t(\xi)=t^{-1/3}\big(h(\xi t^{2/3},t)-2t\sqrt{1-\xi^2 t^{-2/3}}\big).
\end{equation}
In~\cite{PS02} it is shown that, in the sense of finite dimensional distributions,
\begin{equation}\label{cvgAiryPNG}
\lim_{t\to\infty} h^{\rm resc}_t(\xi)=\Airy(\xi),
\end{equation}
where $\Airy$ is the \emph{Airy process}. The definition and properties of the Airy process are given in Section~\ref{sect4.3}. This process arises also in the multi-matrix model for GUE random matrices as we will explain in Section~\ref{GUEAiry}.

More recently Borodin and Olshanski showed~\cite{BO04} that the Airy process describes the space-time correlations not only for the space-time cut with constant $t$, but also along any \emph{space-like} (and \emph{light-like}) path in the droplet geometry. They prove convergence of finite-dimensional distributions in the language of Young diagrams. The link with Young diagrams is explained in Section~\ref{SectPNGYoung}. For each point $(u,v)\in\R_+^2$, they consider the random Young diagram $Y(u,v)$ obtained by the RSK correspondence. Then for each space-like path in $\R_+^2$ a Markov chain is constructed, which describes the evolution of the Young diagram $Y$. The case $u+v=t$ is the one of the PNG droplet~\cite{PS02}. The case $u v = \textrm{\emph{constant}}$ corresponds to the terrace-ledge-kink (TLK) model which can be regarded as model for the facet boundary of a crystal in thermal equilibrium~\cite{FPS03,FS03}. For \emph{time-like} paths no result is known. The major difficulty seems to lie in the lack of a Markov property.

\subsection{Flat PNG}
\begin{figure}[t!]
\begin{center}
\psfrag{h}[l]{$h(x,t)$}
\psfrag{x}[c]{$x$}
\includegraphics[height=5cm]{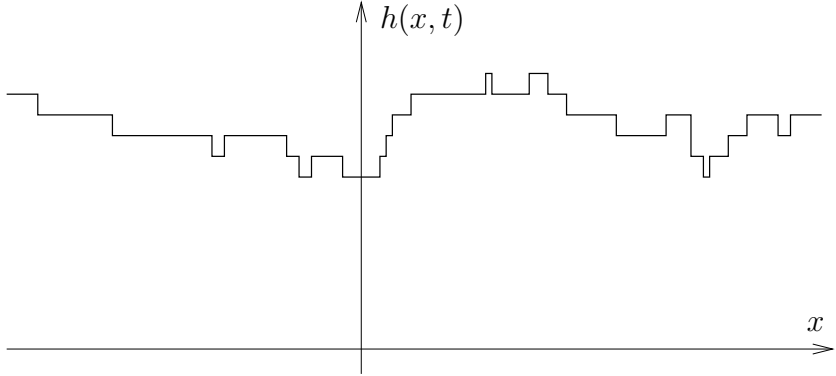}
\caption{A sample of the flat PNG.}\label{FigFlat}
\end{center}
\end{figure}

A flat initial condition, $h(x,0)=0$ for $x\in\R$, and constant density of Poisson points in $\R\times\R_+$ (as before we choose $\varrho=2$) generates the flat PNG geometry, see Figure~\ref{FigFlat}. Since no other constraint is imposed the surface height $h(x,t)$ is statistically translation-invariant, thus we focus on $x=0$.

The fluctuations live on the same scale as for the PNG droplet, namely $t^{1/3}$. However, the limiting distribution is different as shown in~\cite{PS00}, namely
\begin{equation}\label{eqComb14}
\lim_{t\to\infty}\Pb\big(h(0,t)\leq 2t+t^{1/3}2^{-2/3}s\big) = F_1(s),
\end{equation} 
with $F_1$ the GOE Tracy-Widom distribution function~\cite{TW96}. This result follows from a related problem on longest increasing subsequences~\cite{BR99}. The convergence is in distribution and for all moments as in (\ref{2.3}).

In~\cite{Fer04} a first step towards understanding the spatial correlation is made, as we will explain in Sections~\ref{sect4.3} and~\ref{sect5.5}. In a recent preprint, Sasamoto~\cite{Sas05} obtained the spatial correlation for a closely related model. By universality this new process should be the analogue of the Airy process for flat PNG.

\subsection{A geometrical point of view for the PNG model}\label{SectGraph}
Let us illustrate the space-time picture of the PNG model, which is the starting point for the link to other equivalent models described in the next section.

\begin{figure}[t!]
\begin{center}
\psfrag{x}{$x$}
\psfrag{t}{$t$}
\psfrag{t=T}{$t=T$}
\psfrag{h0}[c]{$h=0$}
\psfrag{h1}[c]{$h=1$}
\psfrag{h2}[c]{$h=2$}
\psfrag{h3}[c]{$h=3$}
\includegraphics[height=5cm]{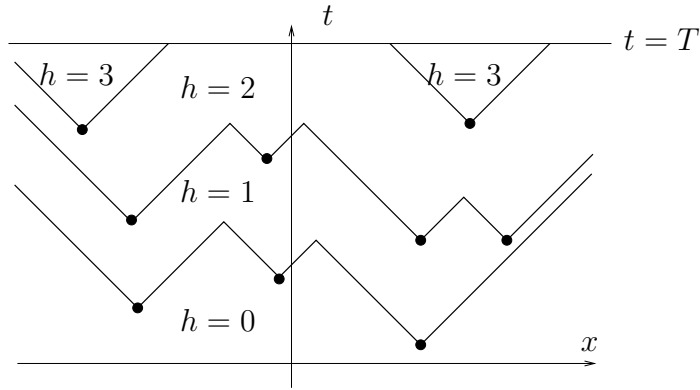}\caption{Graphical construction generating the surface height from the nucleation events (Poisson points).}\label{figRSK2}
\end{center}
\end{figure}

First we consider the case of flat initial conditions, $h(x,0)=0$ for all $x\in\R$. For a given realization of Poisson points, we construct $h(x,t)$ for $t\in [0,T]$, $T>0$ fixed, as follows. One starts plotting the nucleation events in space-time. Then, increasingly in time, one draws the trajectories of the up- and down-steps in space-time. These are light-like paths. When two of these paths meet, as $t$ increases, they stop. This reflects the disappearing of the corresponding up- and down-step.
In this way one divides space-time into regions bounded by piecewise straight lines with slopes $\pm 1$, see Figure~\ref{figRSK2}.
The height $h(x,t)$ is constant in each region and is given by the number of lines crossed by any time-light path from $(x,0)$ to $(x,t)$. For any given $t$, since the Poisson process is a.s.\ locally finite, so is the number of steps of $x\mapsto h(x,t)$, thus $h$ is a.s.\ locally bounded.

In the case that the initial surface profile is not flat, the surface height at some later time $t$ is obtained in a similar way. The only difference is the following. To the lines generated by the Poisson points we need to add additional lines starting from the $t=0$ axis with slope $-1$, resp.\ $+1$, if initially at $x$ there is an up-step, resp.\ a down-step. Then, the number of lines crossed along any time-like paths from $(x,0)$ to $(x,t)$ is the height difference $h(x,t)-h(x,0)$.

\subsection{Other geometries studied for the PNG model}
As explained above, the one-point distribution function of the surface height for both geometries, the PNG droplet and flat PNG, was obtained by identifying the surface height with the length of a longest directed polymer. The directed polymer can be mapped to a longest increasing subsequence in a random permutation (without/with involution), for which Baik and Rains have obtained the asymptotics~\cite{BR99,BR99b} by using Riemann-Hilbert techniques. For the PNG droplet, the joint-distribution of the height profile is obtained in~\cite{PS02} using a multilayer generalization of the PNG as will be explained in Section~\ref{LineEnsembles}. The multilayer technique has been motivated by the work of Johansson on the Aztec diamond~\cite{Jo02b,Jo03}.

Sasamoto and Imamura studied the (discrete) half-droplet PNG geometry, which consists in allowing nucleations only in the region $x\in [0,t]$~\cite{SI03}. They prove that the rescaled height is GUE distributed away from $x=0$ and that there is a transition to GSE at $x=0$. If extra nucleations are added at the origin with intensity $\gamma\geq 0$, the distribution above $x=0$ has a transition at $\gamma=1$. For $\gamma<1$ it is still GSE, for $\gamma=1$ it is GOE distributed, and for $\gamma>1$ the fluctuations become Gaussian, because the contribution of the nucleations at the origin dominates. The asymptotic one-point distributions at the origin follow from~\cite{BR99,BR99b}, too.

A modification of the PNG droplet consists in adding sources at both extremities of the droplet. Extra nucleations with fixed line density $\alpha_+$ and $\alpha_-$ are independently added along the boundary of the forward light cone of the origin, i.e., in $(x,t)$ such that $|x|=t$. This model was used in~\cite{PS00} to describe stationary PNG growth. Baik and Rains~\cite{BR00} obtained detailed results for the asymptotic distributions which we describe briefly. For $\alpha_\pm$ small, the effects coming from the edges are small and the fluctuations are still GUE distributed. On the other hand, if $\alpha_+>1$ or $\alpha_->1$, the boundary effects are dominant and the fluctuations become Gaussian. The cases where $\alpha_+=1$ and/or $\alpha_-=1$ are also studied and other statistics arise. Of particular interest is when $\alpha_+ \alpha_- =1$ for $1-\alpha_\pm=\Or(t^{-1/3})$, in which case the PNG growth is \emph{stationary} and has a flat limit shape. 
Stationary PNG together with flat PNG and PNG droplet are the three most interesting situations in the context of surface growth.

Johansson in~\cite{Jo00b} pointed out the connection between random matrices and the shape fluctuations in a discrete model of directed polymers, equivalent to the discrete PNG model. A class of corner growth models in discrete space-time (similar to the PNG droplet) is analyzed in~\cite{GTW00,MN04b}. The discrete versions of the different geometries for the PNG model discussed above are studied in a series of papers~\cite{Jo03b,SI03,SI04,SI04b}.

\section{Equivalent models}\label{OtherModels}
In this section we discuss the mapping between models which are equivalent to the PNG model.
We start explaining the directed polymers on Poisson points. The link to the PNG was used in~\cite{PS00} to obtain the first results on the height fluctuations of the PNG model. Then we continue with the longest increasing subsequence problem, Young tableaux, a directed percolation model, the kink-antikink gas, and the Hammersley process.

\subsection{Directed polymers on Poisson points}
Let us define a partial ordering in $\R^2$, $\prec$, as follows. For $y,z\in \R^2$ we say that $y\prec z$ if both coordinates of $y$ are less or equal than those of $z$. Consider a Poisson process with intensity $1$ in $\R^2$. For a given realization of Poisson points, a \emph{directed polymer on Poisson points} starting at $(0,0)$ and ending at $(t,t)$ is a piecewise linear path $\gamma$  connecting $(0,0) \prec q_1 \prec \ldots \prec q_{l(\gamma)} \prec (t,t)$, where $q_i$ are Poisson points. The length $l(\gamma)$ of the directed polymer $\gamma$ is the number of Poisson points visited by $\gamma$. The basic observable of interest is the maximal length of the directed polymers,
\begin{equation}
L(t)=\max_{\gamma} l(\gamma).
\end{equation}
This is called \emph{point-to-point} setting because both initial and final points are fixed. 

A modification of the problem consists in considering the set of directed polymers starting from $(0,0)$ and ending in the segment $U_t=\{(y,z)\in\R_+^2|y+z=2t\}$. This is called the \emph{point-to-line} problem and the maximal length is denoted by $L_\ell(t)$. 

\begin{figure}[t!]
\begin{center}
\psfrag{x}[l]{$x$}
\psfrag{t}[l]{$t$}
\psfrag{h(x,t)}[l]{$h(0,t)$}
\psfrag{(0,0)}[r]{$(0,0)$}
\psfrag{(t1,t1)}[l]{$(t/\sqrt{2},t/\sqrt{2})$}
\includegraphics[height=5cm]{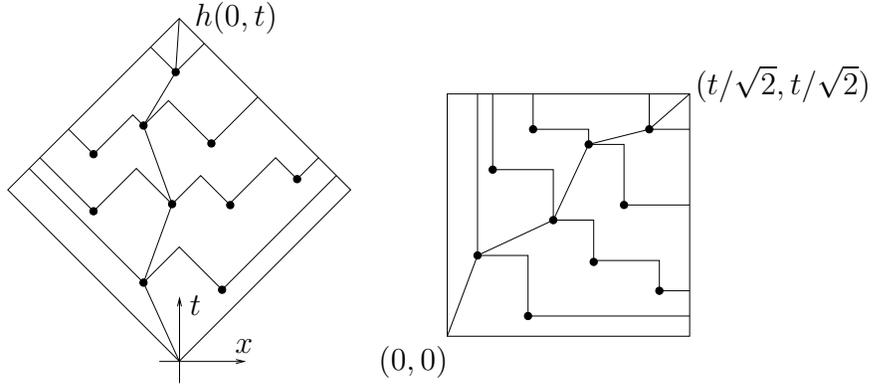}\caption{Height and directed polymers for the droplet geometry}\label{FigDPPNG}
\end{center}
\end{figure}

The link with the PNG model is apparent once we use the graphical point of view explained in Section~\ref{SectGraph}. In fact, for the PNG model, $h(0,t)$ equals the number of lines (up- and down-steps trajectories) crossed by any light-like paths from $(0,0)$ to $(0,t)$. In particular, one considers the paths which cross them at the nucleation points and consist in straight segments between these points. These are the directed polymers introduced above, up to a $\pi/4$ rotation, see Figure~\ref{FigDPPNG}.
Because of the $\pi/4$ rotation, if the density of Poisson points in the PNG model and the directed polymer is the same, then $h(0,t)$ equals $L(t/\sqrt{2})$. To have a nicer formula, we set the density of Poisson points to $2$ for the PNG model and to $1$ for the directed polymers. This implies that $L(t)=h(0,t)$ of the PNG droplet and $L_\ell(t)=h(0,t)$ of the flat PNG, both in distribution. Thus
\begin{equation}\label{eqComb3}
\lim_{t\to\infty}\Pb\big(L(t)\leq 2t + s t^{1/3}\big) = F_2(s),\quad \lim_{t\to\infty}\Pb\big(L_\ell(t)\leq 2t + s 2^{-2/3}t^{1/3}\big) = F_1(s).
\end{equation}

\subsection{Longest increasing subsequences}\label{subsectIncrSubs}
Let $\Perm_N$ denote the permutation group of the set $\{1,\ldots,N\}$. For each permutation $\sigma\in \Perm_N$, the sequence $(\sigma(1),\ldots,\sigma(N))$ has an increasing subsequence of length $k$, $(\sigma(n_1),\ldots,\sigma(n_k))$, if $\sigma(n_1) < \sigma(n_2) < \ldots < \sigma(n_k)$ and $n_1<n_2<\ldots<n_k$.
Denote by $L_N(\sigma)$ the length of the longest increasing subsequences for the permutation $\sigma$. The problem of finding the asymptotic law of $L_N$ for a uniform distribution on $\Perm_N$ is also called Ulam's problem (1961)~\cite{Ul61}. For a review around this problem, see~\cite{AD99}. Baik, Deift and Johansson in the seminal paper~\cite{BDJ99} determined the fluctuation law of $L_N$. They prove
\begin{equation}\label{eqComb5}
\lim_{N\to\infty} \Pb(L_N\leq 2\sqrt{N}+s N^{1/6}) = F_2(s),
\end{equation}
where $F_2$ is the GUE Tracy-Widom distribution function. Compare (\ref{eqComb5}) with (\ref{eqComb3}), the role of $t$ is taken over by $\sqrt{N}$.

(\ref{eqComb5}) is obtained using the Poissonized version of the problem, which is the approach of the problem used by Hammersley~\cite{Ham72}. Instead of fixing the length of the permutations to $N$, one considers the set of permutations $\Perm=\cup_{n\geq 0} \Perm_n$ and assigns the probability $e^{-N} k^N/k!$ that a permutation is in $\Perm_k$. Baik et al.\ first prove that (\ref{eqComb5}) holds for this problem, and then obtain the result via a de-Poissonization method, consisting in bounding from above and below the distribution of $L_N$ in terms of the Poissonized one. In a statistical physics language, the problem with fixed $N$ corresponds to the canonical ensemble, the one with Poisson distributed length to the grand canonical ensemble, and the $N\to\infty$ limit to the thermodynamical limit. It is not surprising that the equivalence of ensembles holds for the observable $L_N$. In fact, in the grand canonical ensemble the typical value of the number of Poisson points is ${\cal O}(\sqrt{N})$ apart from $N$. Thus the correction to $L_N$ is of order $1$, which is vanishing small compared to $N^{1/6}$ as $N\to\infty$.

\begin{figure}
\begin{center}
\psfrag{i}[c]{$i$}
\psfrag{j}[l]{$\sigma(i)$}
\psfrag{1}[c][t]{$1$}
\psfrag{2}[c][t]{$2$}
\psfrag{3}[c][t]{$3$}
\psfrag{4}[c][t]{$4$}
\psfrag{5}[c][t]{$5$}
\psfrag{s}[r]{$\begin{array}{lcr}\sigma&=&(2,3,1,5,4)\\[6pt]
{\cal P}(\sigma)&=&\left(\begin{array}{ccc}1&3&4\\2&5\end{array}\right) \\[18pt]
{\cal Q}(\sigma)&=&\left(\begin{array}{ccc}1&2&4\\3&5\end{array}\right)
\end{array}$}
\includegraphics[height=5cm]{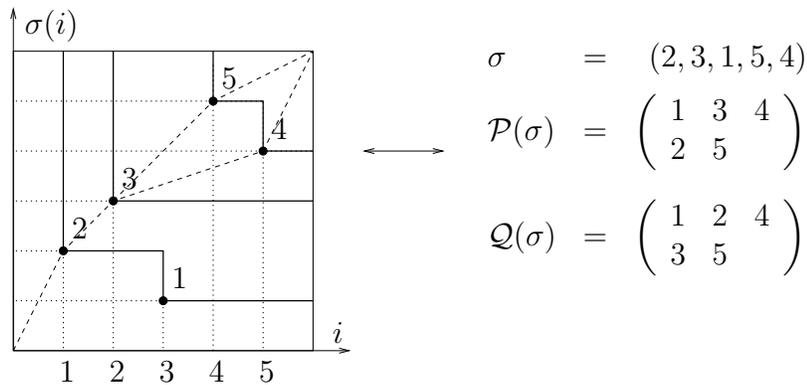}\caption{Longest increasing subsequences and directed polymers. $(2,3,5)$ and $(2,3,4)$ are the two longest increasing subsequences. The directed polymers of maximal length passes through the points labelled by $(2,3,5)$ and $(2,3,4)$.}\label{figPermutation}
\end{center}
\end{figure}

The problem of the longest increasing subsequences in $\Perm_N$ is equivalent to the problem of finding the longest directed polymer from $(0,0)$ to $(t,t)$ when $N$ points are distributed uniformly in the square $[0,t]^2$. The directed polymers on Poisson points is the Poissonized version. In fact, consider a configuration of $N$ points in the square $[0,t]^2$. The length of the longest directed polymer depends only on the order of their projections along both axis. Without changing this order we can put them on $\{1,\ldots,N\}^2$. Thus to each directed polymer there corresponds an increasing subsequence with the same length. Hence the length of the longest directed polymer equals the one of the longest increasing subsequence. See Figure~\ref{figPermutation} for an example.

\subsection{Young tableaux}\label{subsectYoung}
Let $\lambda=(\lambda_1,\lambda_2,\ldots,\lambda_k)$ be a partition of the integer $N$, i.e., satisfying $\lambda_1\geq \lambda_2 \geq \ldots \lambda_k\geq 1$ and $\sum_{i=1}^k\lambda_i=N$. $\lambda$ is represented by a diagram with $k$ rows and with $\lambda_i$ cells for row $i$, called Young diagram. A (standard) Young tableau of \emph{shape} $\lambda$ is a Young diagram, where the cells are filled by the numbers $1,2,\ldots,N$, increasingly in each row and column. The Robinson-Schensted correspondence is a \emph{bijection} between permutations $\sigma\in \Perm_N$ and \emph{pairs} of Young tableaux $({\cal P}(\sigma),{\cal Q}(\sigma))$ with $N$ cells and the same shape. The algorithm leading to $({\cal P}(\sigma),{\cal Q}(\sigma))$ is the following~\cite{Sch61}.\\
\begin{changemargin}{1cm}{1cm}
\noindent One starts with a pair of empty tableaux ${\cal P}$ and ${\cal Q}$ and fills the cells as follows:\\[6pt]
\noindent {\bf ${\cal P}$-tableau}: for $i$ from $1$ to $N$, the number $\sigma(i)$ is always placed via \emph{row-bumping} in the first row, i.e.,\\
a) if $\sigma(i)$ is larger than all numbers in the first row of the ${\cal P}$-tableau, then append to the right of them,\\
b) otherwise put it at the place of the smallest entry in the first row of ${\cal P}$, which is larger than $\sigma(i)$.\\
In case b), the entry which was replaced is now placed via row-bumping in the second row. If an entry is replaced in the second row, then it is placed via row-bumping in the third row, and so on.\\[6pt]
{\bf ${\cal Q}$-tableau}: At each step in the generation of the ${\cal P}$-tableau, its shape is enlarged by one cell. For each $i$ from $1$ to $N$, put the number $i$ at the position where a new cell appeared at step $i$ in the ${\cal P}$-tableau.\\
\end{changemargin}
\noindent As an illustration in Table~\ref{tablePQ} we show  the construction of the Young tableaux for the permutation $\sigma=(2,3,1,5,4)$ of Figure~\ref{figPermutation}.
\tabcolsep=4pt
\begin{table}
\begin{center}
\begin{tabular}{|c|c|c|c|c|c|}
\hline & & & & & \\[-8pt]
$i$ & 1 & 2 & 3 & 4 & 5  \\
\hline & & & & & \\[-8pt]
$\sigma(i)$ & 2 & 3 & 1 & 5 & 4  \\
\hline & & & & & \\[-8pt]
  $\begin{array}{c}
   {\cal P} \\ \\
 \end{array}$ &
 $\begin{array}{c}
   2 \\ \\ 
 \end{array}$ &
 $\begin{array}{cc}
   2 & 3 \\ \\ 
 \end{array}$ &
 $\begin{array}{ccc}
   1 & 3 \\
   2 
 \end{array}$ & 
 $\begin{array}{ccc}
   1 & 3 & 5\\
   2 
 \end{array}$ &
 $\begin{array}{ccc}
   1 & 3 & 4\\
   2 & 5 
  \end{array}$
 \\
  & & & & &  \\[-12pt]
  \hline & & & & &  \\[-8pt]
 $\begin{array}{c}
   {\cal Q} \\ \\
 \end{array}$ &
 $\begin{array}{c}
   1 \\ \\
 \end{array}$ &
 $\begin{array}{cc}
   1 & 2 \\ \\
 \end{array}$ &
 $\begin{array}{cc}
   1 & 2 \\
   3
 \end{array}$ & 
 $\begin{array}{ccc}
   1 & 2 & 4\\
   3
 \end{array}$ &
 $\begin{array}{ccc}
   1 & 2 & 4 \\
   3 & 5 
  \end{array}$ \\[-12pt]
  & & & & &  \\
  \hline
\end{tabular}
\caption{Construction of the Young tableaux ${\cal P}(\sigma)$ and ${\cal Q}(\sigma)$ for the permutation $\sigma=(2,3,1,5,4)$.}
\label{tablePQ}
\end{center}
\end{table}
By construction, the Young tableaux ${\cal P}(\sigma)$ and ${\cal Q}(\sigma)$ have the same shape. Given a permutation $\sigma\in \Perm_N$, one places the points on $\N^2$ with coordinates $(i,\sigma(i))$ and draws the lines as in Figure~\ref{figPermutation}. Notice that the first row of ${\cal P}(\sigma)$ contains precisely the positions of the horizontal lines at abscissa $N+1/2$, likewise the first row of ${\cal Q}(\sigma)$ contains the positions of the vertical lines at ordinate $N+1/2$. This property holds for all permutations which means that~\cite{AD99}
\begin{equation}\label{eq3.4}
L_N(\sigma)=\lambda_1(\sigma).
\end{equation}

Consequently, a way to determine the asymptotic behavior of $L_N$ is by analyzing the length of the first row of Young tableaux~\cite{VK77,LS77}. The measure on the set of partitions of $\{1,\ldots,N\}$, $Y_N$, induced by the uniform measure on $\Perm_N$ via the RS correspondence is the \emph{Plancherel} measure $\textrm{Pl}_N$; let $d_\lambda$ denote the number of Young tableaux of shape $\lambda$, then
\begin{equation}\label{Plancherel}
\textrm{Pl}_N(\lambda)=\frac{d_\lambda^2}{\sum_{\mu\in Y_N}d_\mu^2},\quad \lambda\in Y_N.
\end{equation}

The lengths $\lambda_2(\sigma),\lambda_3(\sigma),\ldots$ also have an interpretation in terms of directed polymers. This is discussed in Section~\ref{SectPNGYoung}.

\subsection{A directed percolation model}
Motivated by the discrete anisotropic directed polymers model of Rajesh and Dhar~\cite{RD98}, one can reformulate the PNG model starting from a directed polymer picture as follows. We take a stack of horizontal planes $\R^2\times \N\subset \R^3$. Neighboring planes are connected by vertical bonds placed randomly. One introduces the percolation cluster $P\subset \R^2\times \N$ as follows. In each plane there is perfect directed percolation, i.e., if $(x,y,k)\in P$ then $\{(x',y',k) | x'\geq x,y'\geq y\}\subset P$. Percolation between adjacent planes occurs through the bonds, i.e., if there is a bond from $(x,y,k)$ to $(x,y,k+1)$, then $(x,y,k+1)\in P$ whenever $(x,y,k)\in P$.

If the position of the bonds between planes are given by independent Poisson processes of intensity $1$, then the cluster spreading from the origin, $(0,0,0)$, corresponds to the point-to-point directed polymer. The height of the cluster at $(x,y)$ is the largest $k$ such that $(x,y,k)\in P$. Thus the height at $(x,y)$ equals, in distribution, the length of the longest directed polymer from $(0,0)$ to $(x,y)$. Similarly, the point-to-line directed polymer corresponds to the cluster spreading from the line $\{(x,y,k) | x+y=0, k=0 \}$. Define the \emph{time} axis as $\{(x,x) | x\in\R_+\}$. Then the PNG dynamics is recovered by slicing the percolating cluster perpendicularly to the time axis.

The percolation cluster arises as the continuum limit of the discrete model of Rajesh and Dhar~\cite{RD98}. It appears also in the limit of large alphabets for a model of sequence aligning~\cite{MN04}.

\subsection{Interacting particle models}
\subsubsection*{The kink-antikink gas}
Bennet et al.~\cite{BBLT81} studied an idealized model for solitons in the sine-Gordon model. There are two types of solitons, called kinks and antikinks. Kinks move to the left with velocity $-1$, antikinks to the right with velocity $1$. The solitons are point-like and do not interact, with the exception that, if a kink and an antikink collide, they annihilate each other. On the other hand, there is a constant uniform rate of production of kink-antikink pairs, which immediately move apart with unit speed. In~\cite{BBLT81} the stationary distribution of the kink-antikink gas is studied on a finite ring and in the thermodynamic limit. From the description of the PNG model it is clear that the kinks and antikinks can be identified with the up- and down-steps of the PNG height profile. Thus any result for the PNG model has a translation to statements for the kink-antikink gas and vice versa, provided initial and boundary conditions match. For example, the height above the origin for flat PNG equals the number of kinks and antikinks that passed through the origin (with vacuum as initial condition).

\subsubsection*{Hammersley process}
In order to tackle the longest increasing subsequence problem, Hammersley~\cite{Ham72} used a particle system to prove that $\lim_{N\to\infty}\E(L_N)/\sqrt{N}$ exists, see also~\cite{AD95,Se96}. There is only one kind of particles on $[0,T]$, which sometimes jump to the left a certain distance, but otherwise are at rest. The position $x=T$ serves as particle reservoir and initially no particles are in $[0,T)$. If a particle jumps, it jumps to the left to a position uniformly chosen in the interval between the particle and its left neighbor (or, for the left-most particle, the origin). The jump rate for each particle is proportional to the distance to its left neighbor. Particles do not cross under the dynamics. Moreover, the space-time points where the particles land form a Poisson point process in $[0,T)\times\R_+$ with uniform density. It is easy to see that for a given realization of Poisson points, the space-time trajectories of the particles correspond, up to a $\pi/4$ rotation, to the light-like lines generated by the nucleations in the PNG model of Figure~\ref{FigDPPNG}. The PNG with sources corresponds to the Hammersley process with sources and sinks~\cite{CG05}.

\section{Description of the PNG via line ensembles}\label{LineEnsembles}
\subsection{Non-intersecting line ensembles}
The surface height of the PNG model at time $T$, $x\mapsto h(x,T)$, does not contain anymore the information of the position of the Poisson points, because whenever two steps collide, information is lost. Therefore the measure induced by the Poisson process on the set of heights is not easy to describe. A way of recording the lost information is to extend the model to a multilayer version. This is achieved using the Robinson-Schensted-Knuth (RSK) construction~\cite{Vie77}. 

Instead of a single line $x\mapsto h(x,t)$ evolving in time according to the PNG dynamics, one considers a set of lines $\{x\mapsto h_\ell(x,t) | \ell \leq 0\}$, with the identification $h_0\equiv h$. The initial condition is $h_\ell(x,0)=\ell$, $x\in \R$, $\ell\leq 0$. The deterministic dynamics is identical for all the lines. Only the stochastic part differs as follows. The first line follows the PNG dynamics as explained in Section~\ref{PNGmodel}. If at time $t$ an up- and a down-step annihilate at position $x$, one records the information in the second line in the form of a nucleation event at $(x,t)$. This procedure is repeated recursively, that is, the nucleation events in the line $h_{\ell-1}$ correspond to the annihilation events in the line $h_\ell$.

\begin{figure}[t!]
\begin{center}
\psfrag{x}{$x$}
\psfrag{t}{$t$}
\psfrag{t=T}{$t=T$}
\psfrag{1}{$\ell=0$}
\psfrag{2}{$\ell=-1$}
\psfrag{3}{$\ell=-2$}
\psfrag{4}{$\ell=-3$}
\psfrag{j}{$j$}
\psfrag{h1}[r]{$h_{0}$}
\psfrag{h2}[r]{$h_{-1}$}
\psfrag{h3}[r]{$h_{-2}$}
\psfrag{h4}[r]{$h_{-3}$}
\psfrag{h5}[r]{$h_{-4}$}
\includegraphics[width=10cm]{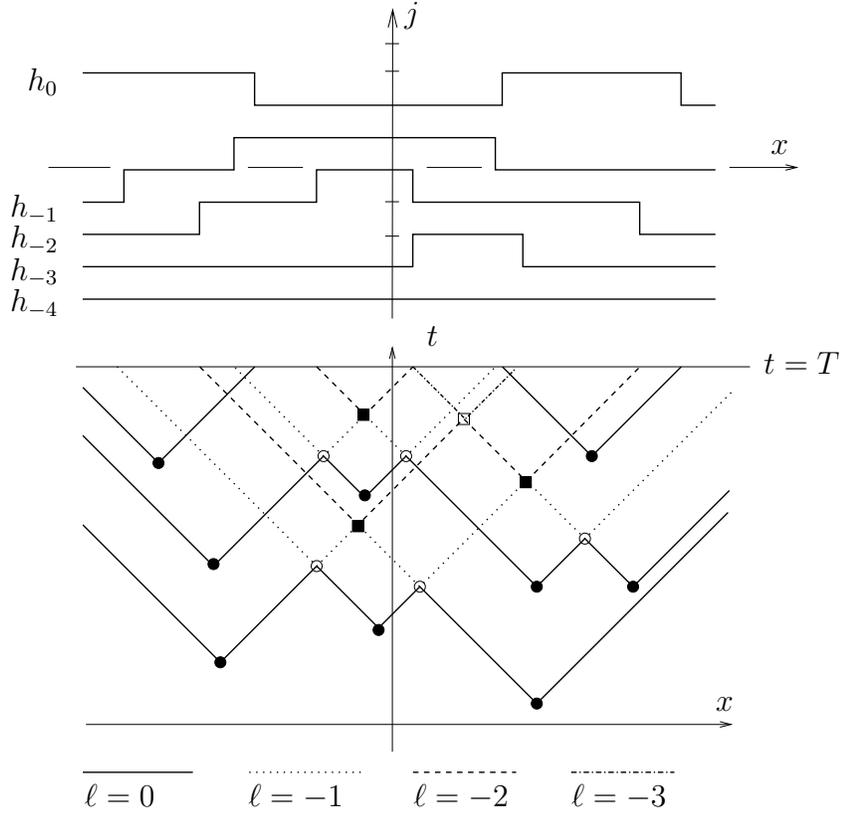}\caption{RSK construction up to time $t=T$. The nucleation events of level $-1$ are the empty dots and the light-like lines of level $2$ are the dotted lines. The corresponding line ensemble is represented above.}\label{figRSK3}
\end{center}
\end{figure}

As for the single line, also for the multilayer version of the PNG model there is a geometric point of view. In space-time, when two light-like lines generated by the Poisson points meet, they become light-like lines for the second level. More generally, when two light-like lines of level $\ell$ meet, they continue as lines of level $\ell-1$. This is illustrated in Figure~\ref{figRSK3}.
By construction, the number of lines of level $\ell$ crossed along any time-like path from $(x,0)$ to $(x,t)$ is greater than the one of level $\ell-1$, for all $\ell\leq 0$. Therefore the set of lines has the non-intersecting property
\begin{equation}\label{eqLines3}
h_\ell(x,t)\geq h_{\ell-1}(x,t)+1
\end{equation}
for all $x\in\R$, $t\geq 0$, $\ell\leq 0$.

\subsubsection*{Non-intersecting line ensemble for $t=T$}
The RSK construction gives us the set of height functions $\{h_\ell, \ell \leq 0\}$. If we want to look at the PNG height at fixed time, say $t=T$, we consider the set of height functions $x\mapsto h_\ell(x,T)$, $\ell\leq 0$. By (\ref{eqLines3}) $\{h_\ell(\cdot,T),\ell\leq 0\}$ is a set of non-intersecting lines with $x\mapsto h_0(x,T)$ the surface profile at time $T$, see Figure~\ref{figRSK3}.

\subsubsection*{Non-intersecting line ensemble for other space-time cuts}
In some situations, as in the work on the flat PNG~\cite{Fer04}, it can be convenient to analyze a line ensemble which corresponds to another space-time cut. Consider a continuous and piecewise differentiable path $\gamma:I\to \R\times[0,T]$, $I\subset \R$ an interval. Then the line ensemble corresponding to $\gamma$, denoted by $\{H_\ell, \ell\leq 0\}$, is given by $H_\ell(s)=h_\ell(\gamma(s))$, $s\in I$, $\ell\leq 0$. It is a non-intersecting line ensemble because of (\ref{eqLines3}).

\subsection{Line ensembles and (real valued) Young tableaux}\label{SectPNGYoung}
\begin{figure}[t!]
\begin{center}
\psfrag{g1}{$\gamma_1$}
\psfrag{g2}{$\gamma_2$}
\psfrag{H02}{$H_0^{(2)}$}
\psfrag{H12}{$H_1^{(2)}$}
\psfrag{H01}{$H_0^{(1)}$}
\psfrag{H11}{$H_1^{(1)}$}
\psfrag{s}[r]{$\sigma=(2,3,1)$}
\psfrag{P}[r]{${\cal P}(\sigma)=\left(\begin{array}{cc}1&3\\2\end{array}\right)$}
\psfrag{Q}[r]{${\cal Q}(\sigma)=\left(\begin{array}{cc}1&2\\3\end{array}\right)$}
\includegraphics[height=5cm]{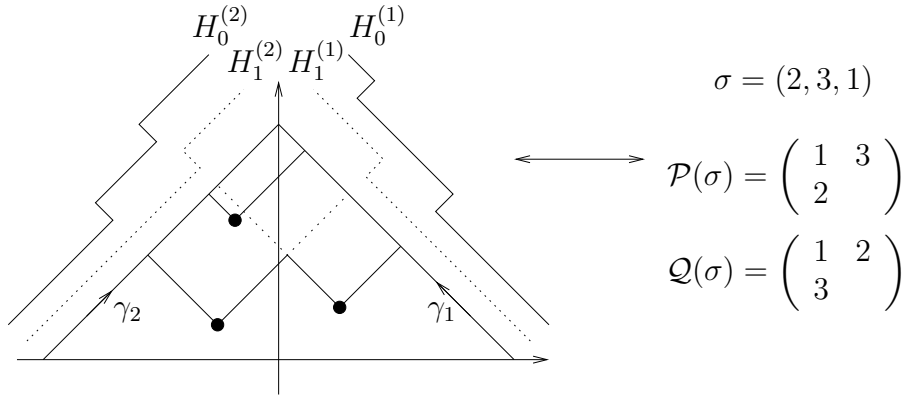}\caption{Young tableaux and line ensembles.}\label{figYoung}
\end{center}
\end{figure}

We now explain the connection between line ensembles and Young tableaux. The height of the PNG surface above $x=0$ at time $t$ depends only on the Poisson points in the backward light cone of $(0,T)$, i.e., on $\triangle_{(0,T)}=\{(x,t)\in\R\times\R_+ \textrm{ s.t. } |x|\leq T-t\}$. Let us consider the two space-time cuts $\gamma_1,\gamma_2$ given by
\begin{eqnarray}
s\mapsto \gamma_1(s)&=& (T-s,s),\quad s\in [0,T],\nonumber \\
s\mapsto \gamma_2(s)&=& (s-T,s),\quad s\in [0,T].
\end{eqnarray}
Denote by $\{H_\ell^{(1)},\ell\leq 0\}$, resp.\ $\{H_\ell^{(2)},\ell\leq 0\}$, the line ensemble along $\gamma_1$, resp.\ $\gamma_2$. From these two line ensembles we construct a pair of Young tableaux $(Y_1,Y_2)$ as follows. Let us start with $Y_1$. Let $0<s_1<s_2<\ldots<T$ be the positions of steps in the line ensemble $\{H_\ell^{(1)},\ell\leq 0\}$. Let $\ell_i$ denote the line in which the step at $s_i$ occurs. Then $Y_k$ has the entry $i$ in row $j$ if the step at $s_i$ happens in the line $H_{1-j}^{(1)}$, i.e., if $\ell_i=1-j$. $Y_2$ is obtained in the same way with step positions along $\gamma_2$. On the other hand we can define a permutation $\sigma\in \Perm_N$, with $N$ the number of Poisson points in $\triangle_{(0,T)}$, by recording the relative positions of the projections of the Poisson points along $\gamma_1$ and $\gamma_2$. More precisely, we set $(y,z)=(t+x,t-x)$ and label the Poisson points such that $z_i\leq z_{i+1}$. Then, $\sigma$ is the permutation such that $y_{\sigma(i)}\leq y_{\sigma(i+1)}$. See Figure~\ref{figYoung} for a simple example. At first glance surprisingly, we have
\begin{equation}
Y_1={\cal P}(\sigma),\quad Y_2={\cal Q}(\sigma).
\end{equation}
But a closer inspection reveals that the multilayer PNG dynamics is just a translation of the algorithm in Section~\ref{subsectYoung}.
In particular, this implies that
\begin{equation}\label{YoungRM}
H_\ell^{(1)}(T)=H_\ell^{(2)}(T)=\lambda_{1-\ell}+\ell
\end{equation}
where $\lambda_0,\lambda_1,\ldots$ are the length of the rows of the Young tableaux.

For the PNG droplet studied in~\cite{PS02}, the measure on the Young tableaux is the Poissonized Plancherel measure. For flat PNG the measure induced by the Poisson points can still be studied if we introduce the symmetric images of the Poisson points with respect to the $t=0$ axis, see~\cite{Fer04} for details. In~\cite{SI03} a version of the PNG droplet is studied where the Poisson points are symmetric with respect to $x=0$. These two symmetries correspond to two different involutions on the corresponding permutations, denoted by $\boxbslash$ and $\boxslash$ in~\cite{BR99}.

\vspace{6pt}
\textit{Length of the rows of Young tableaux.} Given the permutation $\sigma$ as above, the interpretation of $\lambda_i(\sigma)$, $i=1,\ldots,k$, follows from a theorem of Greene~\cite{Gre74}. Let, for $i\leq k$, $a_i(\sigma)$ be the length of the longest subsequence of $\sigma$ consisting of $i$ disjoint increasing subsequences. Greene proves that
\begin{equation}\label{eqComb6}
a_i(\sigma)=\lambda_1+\cdots+\lambda_i.
\end{equation}
In terms of directed polymers, $a_k$ is the maximal sum of the lengths of $k$ non-intersecting directed polymers from $(0,0)$ to $(t,t)$, where non-intersecting means without common Poisson points.

\vspace{6pt}
\textit{Real-valued Young tableaux.} From a different point of view, the line ensembles can be regarded as a generalization of Young tableaux for real valued entries. Consider a configuration of Poisson points with $N$ points in $\triangle_{(0,T)}$ ordered as above. Then one can run the Robinson-Schensted algorithm with the replacements,
\begin{equation}
i \longrightarrow z_i, \quad \sigma(i) \longrightarrow y_i,
\end{equation}
for $i=1,\ldots,N$. In this way one obtains a pair of Young tableaux with $y_i,z_i$ as entries. This pair of real-valued Young tableaux contains exactly the same information as the line ensembles $(H_\ell^{(1)},H_\ell^{(2)})$. More precisely, $y_i$ (resp.\ $z_i$) is in row $j$ of ${\cal P}$ (resp.\ ${\cal Q}$) if there is a step in the line $H_{1-j}^{(1)}$ at $y_i$ (resp.\ in the line $H_{1-j}^{(k)}$ at $z_i$).

\subsection{Point process associated with the line ensemble}\label{sect4.3}
Our goal is to analyze the surface height $h(x,T)$. Therefore it is natural to consider the line ensemble for $t=T$, $\{x\mapsto h_\ell(x,T), \ell\leq 0\}$. Let us define the extended point process $\eta_T$ on $\R\times\Z$ given by
\begin{equation}
\eta_T(x,j)=\left\{\begin{array}{ll} 1,&\textrm{if there is a line passing through }(x,j),\\ 0, &\textrm{otherwise}. \end{array}\right.
\end{equation}
$\eta_T$ is an \emph{extended} point process in the sense that, for any fixed $x\in \R$, $\eta_T(x,\cdot)$ is a point process on $\Z$~\cite{Nev76}. One can interpret the lines as trajectories of fermions in imaginary time (reflecting the non-intersecting condition) where the particles, thus the associated point process, evolve following a given (stochastic) dynamics~\cite{PS02}. We now explain the structure of the point process $\eta_T$ and the edge scaling for the PNG droplet and flat PNG.

\subsubsection*{Point process for PNG droplet: determinantal}
For the PNG droplet it is enough to consider the line ensemble for $x\in [-T,T]$, since for $x\leq -T$ and $x\geq T$ the lines are straight with $h_\ell(\pm T)=\ell$, $\ell \leq 0$. As shown in~\cite{PS02}, the measure on the configurations is the uniform one. Using the transfer matrix method it is shown that $\eta_T$ is an extended \emph{determinantal point process}, which means that the $n$-point correlation functions $\rho^{(n)}(x_1,j_1;\ldots;x_n,j_n)$ can be expressed as a $n\times n$ determinant. For the PNG droplet the kernel is the extended Bessel kernel $B_T(x_1,j_1;x_2,j_2)$, thus
\begin{eqnarray}
\rho^{(n)}(x_1,j_1;\ldots;x_n,j_n)=\det\big(B_T(x_k,j_k;x_l,j_l)\big)_{1\leq k,l \leq n}.
\end{eqnarray}

To analyze the surface one defines the edge scaling of the point process, compare with (\ref{eq2.8}), by
\begin{equation}
\eta_T^{\rm edge}(\tau,s)=T^{1/3}\eta_T(\tau T^{2/3},\lfloor 2T\sqrt{1-\tau^2 T^{-2/3}}+s T^{1/3}\rfloor).
\end{equation}
In the $T\to\infty$ limit, this extended point process converges to the extended determinantal point process on $\R\times\R$ with extended Airy kernel
\begin{equation}\label{Airykernel}
A(\tau_2,s_2;\tau_1,s_1)=\left\{\begin{array}{rl}
\int_{-\infty}^0\dx{\lambda} \, e^{\lambda(\tau_2-\tau_1)} \Ai(s_1-\lambda)\Ai(s_2-\lambda),&\textrm{for }\tau_2\geq \tau_1, \\[6pt]
-\int_0^\infty\dx{\lambda} \, e^{\lambda(\tau_2-\tau_1)} \Ai(s_1-\lambda)\Ai(s_2-\lambda),&\textrm{for }\tau_2<\tau_1.
\end{array}\right.
\end{equation} 
where $\Ai$ is the Airy function as given in~\cite{AS84}. In particular for $\tau_1=\tau_2$ (\ref{Airykernel}) is equal to
\begin{equation}\label{AiryK}
A(s_2;s_1)=\frac{\Ai(s_2)\Aip(s_1)-\Aip(s_2)\Ai(s_1)}{s_2-s_1},
\end{equation}
the classical Airy kernel~\cite{TW94}.

Let $h_T^{\rm resc}$ be the rescaled height as defined in (\ref{eq2.8}). The joint-distribution of $h_T^{\rm resc}$ at positions $\tau_1,\ldots,\tau_n$ can be written as a Fredholm determinant and in the limit of large growth time $T$
\begin{equation}
\lim_{T\to\infty} \Pb\Big(\bigcap_{k=1}^n\{h_T^{\rm resc}(\tau_k) \leq s_k\}\Big) = \det(\Id-f^{1/2} A f^{1/2})_{L^2((\tau_1,\ldots,\tau_n)\times \R)}
\end{equation}
with $f_j(s)=\Theta(s-s_j)$. In this way in~\cite{PS02} the convergence of $h_T^{\rm resc}$ to the Airy process is determined in the sense of finite-dimensional distributions.

\subsubsection*{The Airy process}
The Airy process is defined by its finite-dimensional distribution~\cite{PS02,Jo03}. For given $s_1,\ldots,s_n\in\R$ and $\tau_1<\ldots<\tau_n\in \R$, we define $f$ on $\Lambda=\{\tau_1,\ldots,\tau_n\}\times \R$ by $f(s_j,x)=\chi_{(s_j,\infty)}(x)$. Then
\begin{equation}
\Pb(\Airy(\tau_1)\leq s_1,\ldots,\Airy(\tau_n)\leq s_n) =\det(\mathbbm{1}-f^{1/2}A f^{1/2})_{L^2(\Lambda,\dx^n x)}\nonumber
\end{equation}
with $A$ the integral operator with extended Airy kernel.

The Airy process $\Airy$ was first introduced in~\cite{PS02} in the context of the PNG droplet. There it was shown that $\Airy(t)$ has a version which is almost surely continuous, stationary in $t$, and invariant under time-reversal. Its single time distribution is given by the GUE Tracy-Widom distribution. In particular, for fixed $t$,
\begin{eqnarray}
\Pb(\Airy(t)>y) &\simeq& e^{-y^{3/2}4/3}\quad\textrm{for }y\to\infty,\nonumber\\
\Pb(\Airy(t)<y) &\simeq& e^{-|y|^3/12}\quad\textrm{for }y\to-\infty,
\end{eqnarray}
Thus the Airy process is localized. Define the function $g$ by 
\begin{equation}
\Var(\Airy(t)-\Airy(0)) = g(t).
\end{equation}
From~\cite{PS02} we know that $g$ grows linearly for small $t$ and that the Airy process has long range correlations:
\begin{equation}
g(t)= \left\{\begin{array}{ll}
2t+\Or(t^2)&\textrm{for }|t|\textrm{ small,}\\
g(\infty)-2 t^{-2}+\Or(t^{-4})&\textrm{for }|t|\textrm{ large.}
\end{array}\right.
\end{equation}
with $g(\infty)=1.6264\ldots$. The coefficient $2$ in front of $t^{-2}$ is determined in~\cite{AvM03,Wid03}. The Airy process has been recently investigated and a set of PDE's~\cite{AvM03} and ODE's~\cite{TW03,TW03b} describing it are determined. 

\subsubsection*{Point process for flat PNG: Pfaffian}
The case of the flat PNG is more difficult and, up to now, the analysis shown above is carried out only at fixed position. Because of spatial translation invariance, we can choose the reference position to be $x=0$. We define the edge scaling of the point process as
\begin{equation}
\eta_T^{\rm edge}(s)=2^{-2/3} T^{1/3} \eta_T(0,\lfloor 2T+s 2^{-2/3} T^{1/3}\rfloor).
\end{equation}
The factor $2^{-2/3}$ is only for convenience, compare with (\ref{eqComb14}). It will make the connection with GOE random matrices more transparent.

The structure of this point process is not determinantal as for the PNG droplet. In~\cite{Fer04} we analyze it and prove that it converges weakly to a Pfaffian point process denoted by $\eta^{\rm GOE}$. More precisely, it is shown that for smooth test functions of compact support $f_1,\ldots,f_m$, $m\in \N$,
\begin{equation}
\lim_{T\to\infty} \E\Big(\prod_{k=1}^m \eta^{\rm edge}_T(f_k)\Big) = \E\Big(\prod_{k=1}^m \eta^{\rm GOE}(f_k)\Big).
\end{equation}

This point process appears also in the edge scaling limit of the eigenvalues of GOE random matrices (see Section~\ref{sect5.3}), hence the name $\eta^{\rm GOE}$. $\eta^{\rm GOE}$ is a Pfaffian point process with kernel $K^{\rm GOE}$, i.e., the $n$-point correlation functions $\rho^{(n)}(s_1,\ldots,s_n)$ are given by
\begin{equation}\label{eqCorrFct}
\rho^{(n)}(s_1,\ldots,s_n) = \Pf\big(K^{\rm GOE}(s_i,s_j)\big)_{1\leq k,l \leq n}
\end{equation}
where $K^{\rm GOE}$ is a $2\times 2$ matrix kernel. For an antisymmetric matrix $A$ of even dimensions, one has the identity $\Pf(A)^2=\det(A)$. The entries of  $K^{\rm GOE}$ are given by
\begin{eqnarray}\label{eqGOEkernel}
K_{1,1}^{\rm GOE}(s_1,s_2)
&=&\int_0^{\infty} \dx\lambda \Ai(s_1+\lambda)\Aip(s_2+\lambda)-(s_1\leftrightarrow s_2), \\
K_{1,2}^{\rm GOE}(s_1,s_2) 
&=&\int_0^{\infty} \dx\lambda \Ai(s_1+\lambda)\Ai(s_2+\lambda)+\frac12 \Ai(s_1) \int_0^{\infty}\dx\lambda \Ai(s_2-\lambda), \nonumber\\
K_{2,1}^{\rm GOE}(s_1,s_2) 
&=&-K_{1,2}^{\rm GOE}(s_2,s_1) \nonumber\\
K_{2,2}^{\rm GOE}(s_1,s_2) 
&=&\frac14 \int_0^\infty\dx\lambda\int_\lambda^\infty\dx\mu \Ai(s_1-\lambda)\Ai(s_2-\mu)-(s_1\leftrightarrow s_2), \nonumber
\end{eqnarray}
with $(s_1\leftrightarrow s_2)$ standing for the previous term with $s_1,s_2$ interchanged.

\section{Random Matrices}\label{RandomMatrices}
In this section we introduce the Gaussian ensembles of random matrices, the Tracy-Widom distributions, and the multi-matrix models. The literature on random matrix is large, the standard reference book is \cite{Meh91}, see also the review~\cite{Pas00}.

\subsection{Gaussian ensembles of random matrices}
Let us define the two random matrix ensembles which are linked to the PNG droplet and the flat PNG. 
\subsubsection*{Gaussian unitary ensemble (GUE)}
One defines a Gaussian measure on the set $\Omega$ of $N\times N$ complex Hermitian matrices which is invariant under unitary transformations.  Let $H\in \Omega$, then
\begin{equation}\label{eqGRM}
\Pb(H)\dx H = \frac{1}{Z}\exp(-\Tr(H^2)/2N)\dx H
\end{equation}
where $\dx H=\prod_{i=1}^N\dx H_{i,i} \prod_{1 \leq i < j \leq N}\dx \textrm{Re}(H_{i,j})\dx \textrm{Im}(H_{i,j})$ is the product measure on the independent coefficients of $N$. Here and below $Z$ stands for the proper normalization constant. The scaling factor $1/2N$ in (\ref{eqGRM}) is chosen in order to simplify the comparison with the results on the PNG.

One interesting quantity of random matrices is the distribution of the eigenvalues $\lambda_1,\ldots,\lambda_N$, which is obtained by integrating over the unitary group. The result is
\begin{equation}\label{eqEVGUE}
\Pb_{2,N}(\lambda_1,\ldots,\lambda_N)\dx\lambda_1\cdots\dx\lambda_N= \frac{1}{Z} |\Delta_N(\lambda)|^2 \prod_{j=1}^N e^{-\lambda_j^2/2N}\dx \lambda_j,
\end{equation}
with $\Delta_N(\lambda)$ the Vandermonde determinant
\begin{equation}
\Delta_N(\lambda)=\det(\lambda_i^{j-1})_{i,j=1}^N = \prod_{1\leq i<j\leq N}(\lambda_j-\lambda_i).
\end{equation}

\subsubsection*{Gaussian orthogonal ensemble (GOE)}
In this case the matrices are real symmetric and the measure is invariant under orthogonal transformations. The probability distribution takes the same form as in (\ref{eqGRM}), with the reference measure $\dx H=\prod_{1\leq i \leq j \leq N}\dx H_{i,j}$. The integration over the orthogonal group leads to the distribution of the eigenvalues $\lambda_1,\ldots,\lambda_N$
\begin{equation}\label{eqEVGOE}
\Pb_{1,N}(\lambda_1,\ldots,\lambda_N)\dx\lambda_1\cdots\dx\lambda_N= \frac{1}{Z} |\Delta_N(\lambda)| \prod_{j=1}^N e^{-\lambda_j^2/2N}\dx \lambda_j.
\end{equation}

\subsection{Edge scaling, Tracy-Widom distributions}
One can focus on the largest eigenvalue's distribution when the size of the matrices $N$ is large. The largest eigenvalue, $\lambda_{\max}$, takes value close to $2N$ and its fluctuations are of order $N^{1/3}$~\cite{Moo90,For93,TW94}. Tracy and Widom determined the scaling function of $\lambda_{\max}$ with the following result~\cite{TW94,TW96}, see also their review paper~\cite{TW02}.
Let $F_{\beta,N}(t)=\Pb_{\beta,N}(\lambda_{\max}\leq t)$, then $F_\beta(s)$ defined by
\begin{equation}\label{eqF}
F_{\beta}(s)=\lim_{N\to\infty}F_{\beta,N}\big(2N+s N^{1/3}\big)
\end{equation}
exists for $\beta=1,2$. $F_1$ and $F_2$ are called the GOE and GUE Tracy-Widom distribution respectively. More precisely,
\begin{equation}
F_2(s)=\exp\Big(-\int_{s}^\infty (x-s) q^2(x)\dx x\Big),\quad F_1(s)=\exp\Big(-\frac12 \int_s^\infty q(x)\dx x\Big) F_2(s)^{1/2}
\end{equation}
where $q$ is the unique solution of the Painlev\'e II equation $q''=s q + 2 q^3$ satisfying the asymptotic condition $q(s)\sim \Ai(s)$ for $s\to\infty$.

\subsection{Eigenvalues' point process and its edge scaling}\label{sect5.3}
We now define the point process of the eigenvalues $\lambda_1,\ldots,\lambda_N$.
\subsubsection*{GUE point process: determinantal}
We denote by $\zeta_{N}^{\rm GUE}$ the point process on $\R$ of the GUE eigenvalues $\lambda_1,\ldots,\lambda_N$, i.e.,
\begin{equation}
\zeta_N^{\rm GUE}(\lambda)=\sum_{j=1}^N\delta(\lambda-\lambda_j),\quad \lambda\in\R,
\end{equation}
with the abuse of notation $\delta(\lambda-\lambda_j)$ for the Dirac measure $\delta_{\lambda_i}$. 
$\zeta_N^{\rm GUE}$ is a determinantal point process, see for example Chapter 5 of~\cite{Meh91}, with kernel given by the Hermite kernel
\begin{equation}\label{eqHermite}
K^{\rm H}_N(x,y)=\frac{p_N(x) p_{N-1}(y)- p_{N-1}(x) p_N(y)}{x-y}e^{-(x^2+y^2)/4N},
\end{equation}
where $p_k(x)=(2\pi N)^{-1/4} (2^k k!)^{-1/2} p^{\rm H}_k(x/\sqrt{2N})$ with the (standard) Hermite polynomials $p^{\rm H}_k(x)=e^{x^2} \frac{\dx^k}{\dx x^k}e^{-x^2}$.

The edge scaling of the point process $\zeta_N^{\rm GUE}$ corresponding to (\ref{eqF}) is
\begin{equation}
\eta_N^{\rm GUE}(\xi)= N^{1/3} \zeta_N^{\rm GUE}(2N+\xi N^{1/3}).
\end{equation}
The limit point process
\begin{equation}
\lim_{N\to\infty}\eta_N^{\rm GUE} = \eta^{\rm GUE}
\end{equation}
is well defined. It is the determinantal point process with Airy kernel (\ref{AiryK}). The GUE Tracy-Widom distribution $F_2(s)$ can also be written as a Fredholm determinant
\begin{equation}\label{eqF2}
F_2(s) = \det(\Id-A)_{L^2((s,\infty),\dx x)}
\end{equation}
with $A$ the integral operator with Airy kernel. For more information on determinantal point processes, see for example~\cite{Sos00}.

\subsubsection*{GOE point process: Pfaffian}
In the same way as for GUE we define the point process of the GOE eigenvalues $\zeta_{N}^{\rm GOE}$. For even $N$, $\zeta_{N}^{\rm GOE}$ is a Pfaffian point process, see for example Chapter 6 of~\cite{Meh91}. For more recent developments on Pfaffian point processes see~\cite{Ra00,Sos03}. The edge rescaled point process is then given by
\begin{equation}\label{FlatPNGeq1.7}
\eta_N^{{\rm GOE}}(\xi)= N^{1/3} \zeta_N^{{\rm GOE}}(2N+\xi N^{1/3}).
\end{equation}
In the $N\to\infty$ limit, the point process $\eta_N^{\rm GOE}$ converges~\cite{TW96} to the point process $\eta^{\rm GOE}$ which has correlation functions given by (\ref{eqCorrFct}). One consequence is that the distribution of the largest eigenvalue can be written as a Fredholm Pfaffian (see Chapter 8 of \cite{Ra00}) or Fredholm determinant on the measurable space $((s,\infty),\dx x)$
\begin{equation}
F_1(s)=\Pf(J-K^{GOE})=\sqrt{\det(\Id+J K^{GOE})}
\end{equation}
with $J(x,y)=\delta_{x,y}\left(\begin{array}{cc} 0&1\\-1&0\end{array}\right)$. An interpretation of the Fredholm determinant as the one on the operator with integral kernel $K^{\rm GOE}$ needs some care as pointed out by Tracy and Widom in~\cite{TW04}, where they actually prove that the finite-$N$ Fredholm determinant converges to $F_1(s)$.

The GOE kernel being a $2\times 2$ matrix is not uniquely defined. In fact using $\Pf(X^t K X)=\det(X)\Pf(K)$~\cite{Ste90} one obtains a family of kernels $K$ leading to the same correlation functions. For example, the kernels reported in~\cite{FNH99,SI03} differ slightly from (\ref{eqGOEkernel}), but they are equivalent since they yield the same point process.

Very recently a new formula for the GOE Tracy-Widom distribution $F_1$ was established. Let $B(s)$ be the operator with kernel $B(s)(x,y)=\Ai(x+y+s)$. It then holds
\begin{equation}\label{eqF1}
F_1(s)=\det(\Id-B(s))
\end{equation}
where the Fredholm determinant is on $L^2(\R_+)$. This is proven in~\cite{FS05b} starting from the work on the flat KPZ growth~\cite{Sas05}.

\subsection{Multimatrix model: Dyson's Brownian motion}\label{sect5.6}
Dyson~\cite{Dys62} noticed that the distribution of eigenvalues (\ref{eqEVGUE}) and  (\ref{eqEVGOE}) is identical to the \emph{equilibrium} probability distribution of the \emph{positions} of $N$ point charges, free to move in $\R$ under the forces deriving from the potential $U$ at inverse temperature $\beta$, with
\begin{equation}
U(x_1,\ldots,x_N)= -\sum_{1\leq i < j \leq N} \ln|x_i-x_j|+\frac{1}{2N\beta}\sum_{i=1}^N x_i^2.
\end{equation}
In the attempt to interpret the Coulomb gas as a dynamical system Dyson considered the positions of the particles in Brownian motion subjected to the interaction forces $-\nabla U$ and a frictional force $f$ (which fixes the rate of diffusion, or equivalently, the time scale). He showed that in terms of random matrices, this is equivalent to the evolution of the eigenvalues when the independent coefficients of the matrix $H=H(t)$ evolve as independent Ornstein-Uhlenbeck processes given by
\begin{equation}\label{eqDetProc17}
P(H(t)=H|H(0)=H_0)\dx H=\frac{1}{Z}
\exp\left(-\frac{\Tr(H-q H_0)^2}{2N(1-q^2)}\right)\dx H
\end{equation}
with $q=\exp(-t/2N)$. The evolution of the eigenvalues satisfies the set of stochastic differential equation
\begin{equation}\label{eq2.9}
\dx \lambda_j(t) = \bigg(-\frac{1}{2N} \lambda_j(t) + \frac{\beta}{2} \sum^N_{\begin{subarray}{l}i=1,\\i\neq j\end{subarray}}
\frac{1}{\lambda_j(t)- \lambda_i(t)} \bigg)\dx t + \dx b_j(t)\,, \quad
j=1,...,N,
\end{equation}
with $\{b_j(t),\, j=1,...,N\}$ a collection of $N$ independent standard Brownian motions. The parameter $\beta$ is $1$ for GOE and $2$ for GUE. We refer to the \emph{stationary process of (\ref{eq2.9}) as Dyson's Brownian motion} (for the eigenvalues). Note that for $\beta\geq 1$ the process is well defined because there is no crossing of the eigenvalues, as proved by Rogers and Shi~\cite{RS93}.

\subsection{GUE extended point process}\label{GUEAiry}
The GUE eigenvalues $\{\lambda_i(t),i=1,\ldots,N\}$ define an extended point process on $\R_+\times \R$,
\begin{equation}
\zeta_N^{\rm GUE}(t,\lambda)=\sum_{j=1}^N \delta(\lambda-\lambda_j(t)).
\end{equation}
It is a determinantal point process~\cite{EM97} with kernel given by the extended Hermite kernel\footnote{According to~\cite{TW03b} this was already described in a MSRI lecture (2002) by Kurt Johansson. It can be derived for example using Theorem 1.7 of~\cite{Jo03b}, details can be found for example in Appendix A.6 of~\cite{FerPhD}.},
\begin{equation}\label{eqPP2}
K_N^H(t_2,\lambda_2;t_1,\lambda_1)=\left\{\begin{array}{ll}\sum_{k=-N}^{-1} e^{k (t_2-t_1)/2N} p_k(\lambda_1)p_k(\lambda_2) e^{-(\lambda_1^2+\lambda_2^2)/4N},& t_2 \geq t_1, \\[6pt]
-\sum_{k=0}^\infty e^{k (t_2-t_1)/2N} p_k(\lambda_1)p_k(\lambda_2) e^{-(\lambda_1^2+\lambda_2^2)/4N},& t_2<t_1,
\end{array}\right.
\end{equation}
where $p_k(x)=p^{\rm H}_{N+k}(x/\sqrt{2N}) (\sqrt{2\pi N} 2^k k!)^{-1/2}$ and $p^{\rm H}_k(x)$ the standard Hermite polynomials. 

The GUE Dyson's Brownian motion is stationary and the edge scaling is the following. Let $t_i=2 \tau_i N^{2/3}$, $\lambda_i=2N+s_i N^{1/3}$, $i=1,2$. In the limit of large $N$, $K_N^{\rm H}$ converges in the edge scaling limit to the extended Airy kernel
\begin{equation}\label{eqPP8}
\lim_{N\to\infty} N^{1/3} K^{\rm H}_N(2 \tau_2 N^{2/3},2N+s_2 N^{1/3};
2 \tau_1 N^{2/3},2N+s_1 N^{1/3}) = A(\tau_2,s_2;\tau_1,s_1).
\end{equation}
This convergence is uniform for $s_1,s_2\in [a,\infty)$ for any fixed $a\in\R$, see for example Appendix A.7 of~\cite{FerPhD}. Moreover, since it has (super-)exponential decay for large $s_1,s_2\to\infty$, one deduces that the largest eigenvalue converges to the Airy process in the sense of finite-dimensional distributions,
\begin{equation}\label{cvgAiryGUE}
\lim_{N\to\infty}N^{-1/3}(\lambda_{\max}(2\tau N^{2/3})-2N)=\Airy(\tau).
\end{equation}

\section{PNG model, Young tableaux, and random matrices}\label{Connections}

\subsection{PNG model}\label{sect5.5}
Although the PNG model and the random matrix ensembles describe systems completely different, we can associate some point process via the multilayer extension and the integration over the symmetry groups, respectively. The point processes happen to have the same mathematical structure:
\begin{itemize}
\item[(a)]determinantal point processes for PNG droplet and GUE random matrices,
\item[(b)]Pfaffian point processes for flat PNG and GOE random matrices.
\end{itemize}
Moreover we have seen that, in the appropriate scaling limit, they converge to the same limit objects. Thus the distribution of the largest eigenvalue and the height of the PNG at fixed position have the same fluctuations in the asymptotic limit.

One natural question is whether such a connection still exists at the level of joint-distribution (extended point processes). The extended point process for the PNG model is naturally defined from the line ensemble. As explained in Section~\ref{sect5.6} one defines the so-called Dyson's Brownian motion (multi-matrix ensembles) and associates to it an extended point process. For GUE random matrices it turns out that the extended point process has, in the limit of large matrix dimension $N$, the extended Airy kernel, the same as for the PNG droplet. As a consequence, the (droplet) surface height and the evolution of the largest (GUE) eigenvalue are described by the same process in the asymptotic limit~\cite{FPS03b}: the Airy process.

For GOE random matrices the answer is not known. It is conjectured that the surface height of the flat PNG and the evolution of the largest (GOE) eigenvalue converge to the same limit process. The result of Ferrari~\cite{Fer04} makes this conjecture more plausible. In fact, we now know that, not only $h(0,T)$ in the limit $T\to\infty$ and properly rescaled is GOE Tracy-Widom distributed, but also that the complete point process $\eta_T$ converges to the edge scaling of Dyson's Brownian motion with $\beta=1$ for fixed time. For $\beta=1$ Dyson's Brownian motion this structure has not been unraveled and the question remains open. Very recently, Sasamoto~\cite{Sas05} obtained a process in the context of the totally asymmetric exclusion process, which, by universality, should describe also the surface height of the flat PNG.

From the point of view of multi-matrices, the difficulty is the fact that the Harish-Chandra/Itzykson-Zuber formula doesn't have a particularly nice analogue for symmetric matrices. From the point of view of the SDE (\ref{eq2.9}) the problem lies in the different factor in front of the drift term, which seems to make it impossible to regard the process as Doob transforms of $N$ independent processes~\cite{Koe04}.

There are other models with the same mathematical structure and showing the same limit behaviour (and scaling functions): the 3D-Ising corner~\cite{OR01,FS03} and the Aztec diamond~\cite{Jo02b,Jo03} which are linked via domino tilings, the vicious random walks and the non-colliding Brownian particles, see for example~\cite{NKT02,KNT03}, and the totally asymmetric exclusion process~\cite{FS05a,Sas05}. Some of the problems presented above can be solved by using orthogonal polynomials point of view instead of the more geometric line ensembles, see for example the review~\cite{Koe04}. In this context the relevant processes are the Schur process~\cite{OR01} and its Pfaffian analogue~\cite{SI04,RB04}.

\subsection{Young tableaux}\label{RMYoung}
In Section~\ref{YoungRM} we described the connection between the length of the row of the Young tableaux and the point process of the multilayer PNG. Above we explained the connection between the multilayer PNG and the eigenvalues of random matrices. Putting the two arguments together it follows that there is a connection between Young tableaux and random matrices. After having seen that the length of the first row of Young tableaux under the Plancherel measure (\ref{Plancherel}) and the largest eigenvalue have the same statistics in the asymptotic limit, it was conjectured by Baik, Deift, and Johansson that the connection would extend to the top rows, respectively top eigenvalues. They proved it for the second row in~\cite{BDJ99b} and the first proof for all top rows is due to Okounkov~\cite{Ok99}. Other proofs are in~\cite{Jo01,Boo00}. In the work on flat PNG~\cite{Fer04} the point process with symmetric images was studied, which corresponds to the involution $\boxbslash$ for permutations. In this case, the measure on the Young tableaux becomes $\textrm{Pl}_N^{(1)}(\lambda)=d_\lambda$, $\lambda \in Y_N^{(1)}$, $Y_N^{(1)}$ being the restriction of $Y_N$ to the Young tableaux with $\lambda_i$ \emph{even} for all $i$. From the proof for the flat PNG~\cite{Fer04} it follows that the top rows of $Y_N^{(1)}$ and the top eigenvalues of GOE have the same limit joint-statistics.

We first learned of the connection between the Young tableaux and the random matrix ensembles for the GOE and GSE random matrix ensembles by Sasamoto~\cite{Sas04}, where he deduced the connection starting from the PNG half-droplet~\cite{SI03}. For the GOE case the restriction that all $\lambda_i$ have to be even is not required. This result is also obtained by Forrester, Nagao, and Rains in~\cite{FNR05} using an approach with orthogonal polynomials.

%\newcommand{\bibliodir}[1]{../../Biblio/#1}
%\bibliographystyle{\bibliodir{patplain}}
%\bibliography{\bibliodir{Biblio}}

\end{document}